\documentstyle[aps]{revtex}
\input{epsf}

\begin{document}
\bibliographystyle{prsty}
\title{Dirac fields in the background of a
magnetic flux string \\ and spectral boundary conditions}%
\author{C. G. Beneventano\thanks{Fellow Fomec
(Fondo para el Mejoramiento de la Ense\~{n}anza de la
Ciencia)-U.N.L.P.}, M. De Francia\thanks{U.N.L.P.} and E. M.
Santangelo\thanks{Member of CONICET(Consejo Nacional de
Investigaciones Cient\'{\i}ficas y T\'{e}cnicas) and U.N.L.P.} }
\address{Departamento de F\'{\i}sica -
Facultad de Ciencias Exactas -
Universidad Nacional de La Plata \\
C.C. 67 (1900) La Plata - Argentina}%

%


\maketitle
\begin{abstract}
We study the problem of a Dirac field in the background of an
Aharonov-Bohm flux string. We exclude the origin by imposing spectral
boundary conditions at a finite radius then shrinked to zero. Thus, we
obtain a
behaviour of the eigenfunctions which is compatible with
the self-adjointness of the radial Hamiltonian  and the invariance under
integer translations
of the reduced
flux.

After confining the theory to a finite region, we check the consistency
with the index theorem, and evaluate its vacuum fermionic number and
Casimir energy. \end{abstract}

\section{Introduction}
\label{section-1}

The consideration of Bohm-Aharonov \cite{Bohm} scenarios is
relevant to different physical problems, mainly 2+1-dimensional
models in superconductivity and particle theory. The inclusion of
the spin is an aspect to which much attention has been devoted.
Many authors have studied this point in connection with the
interaction of cosmic strings with matter
\cite{AlRuWil1989,Russel1988,AlWil1989,Hagen,Flekkoy}.

We will treat the $3+1$-dimensional problem of a Dirac field in
the presence of a flux string. Due to the symmetry of the
background field, the original four dimensional Dirac Hamiltonian
can be written in a block-diagonal form, thus leading to two
$2+1$-dimensional problems.

In this context, after developing the eigenfunctions in a convenient angular basis,
 the need to consider self-adjoint extensions of
the radial Dirac Hamiltonian was recognized
\cite{GerbertJackiw,Gerbert}. A one-parameter family of boundary
conditions at the origin was then shown to arise, which amounts to
establishing a relationship between components of the spinor
rather than asking for the simultaneous finiteness of both of
them. However, it was shown in \cite{manuel} that only two values
of the extension parameter correspond to the presence of a Dirac
delta magnetic field at the origin.

In references \cite{AlRuWil1989,Hagen,Flekkoy}, one of this
possible boundary conditions was obtained, starting from a model
in which the continuity of both components of the Dirac spinor is
imposed at finite radius and the zero-radius limit is taken. As
pointed out in
\cite{AlRuWil1989,Sitenko1996gd,sitenko97,sitenko96,sitenkohep97},when
this boundary condition is imposed at the origin, the invariance
under integer translations of the reduced magnetic flux (or,
equivalently, large gauge transformations) is lost.

In this paper, we adopt the view that the origin is an excluded
point. The plane is thus a punctured one, which has the topology
of a cylinder, and the aforementioned singular gauge
transformations must constitute an invariance of the theory. It is
in this spirit that, after devoting Sec.~\ref{section-2} to
setting the problem, in Sec.~\ref{section-3} we impose on the
Dirac fields spectral boundary conditions of the Atiyah- Patodi-
Singer (APS) type \cite{aps,aps2,aps3,spectral}, taken as in
reference \cite{Zhong,Ninomiya}, at a finite radius.
 We show that, when the zero
radius limit is taken one then obtains, at the origin, other among
the boundary conditions compatible with self-adjointness, which
present a minimal divergence behaviour
\cite{sitenko97,sitenko96,Sitenko1996gd}. They respect the above
mentioned periodicity of the flux and are, at the same time,
compatible with the presence of a
 Dirac delta magnetic field at the origin.

From Sec.~\ref{section-4} on, we confine the whole system within a
punctured circle of radius R, imposing at the external boundary
spectral conditions complementary to those satisfied at the
position of the flux tube. We discuss
 the problem of zero modes and
the compatibility with APS index theorem for manifolds with boundaries.

In Sec.~\ref{section-5}, we determine the energy spectrum and
study the fermionic number and Casimir energy of the confined
system. This last calculation is performed in a $\zeta$ function
approach, following the lines of reference \cite{Leseduarte1996}.

Finally, Sec.~\ref{section-6} contains some comments and
conclusions.

\section{Setting of the problem}
\label{section-2}

We study the Dirac equation for a massless particle in four
dimensional Minkowski space.
\begin{equation}
\left( i \not\!\partial - \not\!\! A \right)\Psi = 0 \label{ec-1}
\end{equation}
in the presence of a flux tube located at the origin, i.e.,
\begin{equation}
\vec{H} = \vec{\nabla} \wedge \vec{A} = \frac{\kappa}{r} \delta
(r) \check{e}_z
\end{equation}
where $\kappa=\frac{\Phi}{2\pi}$ is the reduced flux.

As the gauge potential is $z-$independent, equation (\ref{ec-1})
can be decoupled into two uncoupled two-component equations
\cite{Hagen},by choosing:
\begin{equation}
\gamma^0 = \left(
\begin{array}{cc}
  \sigma_3 & 0 \\
  0 & \sigma_3
\end{array}
\right) \quad \gamma^1 = \left(
\begin{array}{cc}
  i \sigma_2 & 0 \\
  0 & i \sigma_2
\end{array}
\right) \quad \gamma^2 = \left(
\begin{array}{cc}
  -i\sigma_1 & 0 \\
  0 & i\sigma_1
\end{array}
\right) \quad \gamma^3 = \left(
\begin{array}{cc}
  0 & i \sigma_1 \\
  i \sigma_1 & 0
\end{array}
\right)
\end{equation}

In order to avoid singularities, we will consider that only
$r>r_0$ is accessible, and take the limit $r_0\rightarrow 0$,which
is equivalent to having a punctured plane with the removed point
corresponding to the string position.

By taking
\begin{equation}
A_z=0\qquad A_r =0 \qquad A_\theta =\frac{\kappa}{r}\qquad ,{\rm
for}\quad r>r_0
\end{equation}
the Hamiltonian can be seen to be block-diagonal
 \begin{equation}
 H=\left(\begin{array}{cc}
   H_+ & 0 \\
   0 & H_- \
 \end{array}\right)
 \end{equation}
where the two-by-two blocks are given by
\begin{equation}
H_\pm =\left(
\begin{array}{cc}
  0 & i e^{\mp i \theta} \left(\partial_r \pm {\rm B}\right)\\
   -i e^{\pm i \theta} \left(-\partial_r \pm {\rm B}\right) & 0
\end{array}
\right)
\end{equation}
with
\begin{equation}
{\rm B} = -\frac{i}{r} \partial_\theta -\frac{\kappa}{r}
\end{equation}

It should be noticed that these two ``polarizations", which we
will label with
 $s=\pm1$,
correspond to the two inequivalent choices for the gamma matrices
in 2+1 dimensions \cite{Flekkoy}.

From now on, we will be working with $s=1$ (the case $s=-1$ can be
studied in a similar way, and explicit reference will be made to
it whenever necessary).

In this case, we can write:
\begin{equation}
H_+ =\left(
\begin{array}{cc}
  0 & L^\dag \\
  L & 0
\end{array}
\right) \qquad , {\rm with} \quad L=-i e^{i \theta}
\left(-\partial_r +{\rm B}\right) \qquad L^\dag=i e^{-i \theta}
\left(\partial_r +{\rm B}\right)
\end{equation}

and its eigenfunctions:
\begin{equation}
\Psi_E = \left(\begin{array}{c}
  \varphi_E\left(r,\theta\right) \\
  \chi_E\left(r,\theta\right)
\end{array}
\right) \qquad,\quad{\rm satisfy:}\quad
\begin{array}{c}
  L \varphi_E = E \chi_E \\
  L^\dag \chi_E = E \varphi_E
\end{array}
\label{eq-9}
\end{equation}

Now, the two components in $\Psi_E$  have different $\theta$
dependence. In order to make this fact explicit, and to discuss
boundary conditions at $r=r_0$, we introduce \cite{Zhong,Ninomiya}
\begin{equation}
\Psi_E = \frac{1}{\sqrt{r}} \left(
\begin{array}{c}
  e^{-i\frac{\theta}{2}} \varphi_{1E} \left(r,\theta\right)\\
  e^{i\frac{\theta}{2}} \chi_{1E} \left(r,\theta\right)
\end{array}
\right)
\end{equation}
and
\begin{equation}
L_1 =-\partial_r + {\rm B} \qquad L_1^\dag =\partial_r + {\rm B}
\qquad,
\end{equation}
so that
\begin{equation}
L = -i \frac{ e^{i\frac{\theta}{2}}} {\sqrt{r}} L_1
e^{i\frac{\theta}{2}}\sqrt{r}
 \qquad
L^\dag = i \frac{e^{-i\frac{\theta}{2}}} {\sqrt{r}} L_1^\dag
e^{-i\frac{\theta}{2}} \sqrt{r}
\end{equation}
and
\begin{equation}
\begin{array}{l}
  L_1 \varphi_{1E} = i E \chi_{1E} \\
  L_1^\dag \chi_{1E} =-i E \varphi_{1E}
\end{array}
\label{eq-13}
\end{equation}

We expand $\varphi_{1E}$ and $\chi_{1E}$ in terms of
eigenfunctions of B
\begin{equation}
{\rm B}\,e_n = \lambda_n e_n
\end{equation}
which are of the form:
\begin{equation}
e_n = e^{i\left(n+\frac{1}{2}\right)\theta} \qquad,\quad {\rm
with} \qquad
\lambda_n\left(r\right)=\frac{n+\frac{1}{2}-\kappa}{r} \quad,\quad
n\,\epsilon\, Z
\end{equation}
once the condition has been imposed that $\varphi_E$ and $\chi_E$
in equation (\ref{eq-9}) are single-valued in $\theta$.

Thus,we have
\begin{equation}
\begin{array}{l}
  \varphi_{1E}\left(r,\theta\right) =\sum_{n=-\infty}^\infty
  f_n\left(r\right) e^{i \left(n+\frac{1}{2}\right)\theta} \\
  \chi_{1E}\left(r,\theta\right) =\sum_{n=-\infty}^\infty g_n\left(r\right)
  e^{i\left(n+\frac{1}{2}\right)\theta}
\end{array}
\label{eq-16}
\end{equation}

Replacing in (\ref{eq-13}) we obtain, for noninteger $\kappa$
($\kappa=k+\alpha$, with $k$ the integer part of $\kappa$ and
$\alpha$ its fractionary part)

\begin{equation}
\begin{array}{l}
  \varphi_{1E}\left(r,\theta\right) = \sqrt{r} \sum_{n=-\infty}^\infty
  \left(A_n\,J_{n-\kappa}\left(|E|r\right)+
  B_n\,J_{\kappa-n}\left(|E|r\right)\right)
  e^{i \left(n+\frac{1}{2}\right)\theta} \\
  \chi_{1E}\left(r,\theta\right) =-i\frac{|E|}{E}\sqrt{r}
  \sum_{n=-\infty}^\infty
  \left(A_n\,J_{n+1-\kappa}\left(|E|r\right)-
  B_n\,J_{\kappa-n-1}\left(|E|r\right)\right)
  e^{i\left(n+\frac{1}{2}\right)\theta}
\end{array}
\label{eq-17}
\end{equation}
and
\begin{equation}
\Psi_E\left(r,\theta\right) = \left(
\begin{array}{l}
 \sum_{n=-\infty}^\infty \left(A_n\,J_{n-\kappa}\left(|E|r\right)+
  B_n\,J_{\kappa-n}\left(|E|r\right)\right)
  e^{i n\theta} \\
  \sum_{n=-\infty}^\infty -i\frac{|E|}{E}\left(A_n\,J_{n+1-\kappa}\left(|E|r\right)-
  B_n\,J_{\kappa-n-1}\left(|E|r\right)\right)
  e^{i \left(n+1\right)\theta}
\end{array}
\right) \label{eq-18}
\end{equation}

(Of course, for integer $\kappa$, a linear combination of Bessel and
 Neumann functions must be taken).

 Finally, for $s=-1$, the upper and lower components of $\Psi_E$
 interchange, and $E\rightarrow-E$.

\section{Boundary conditions at the origin}
\label{section-3}

As is well known \cite{AlRuWil1989,GerbertJackiw,Gerbert}, the
radial Dirac Hamiltonian in the background of an Aharonov-Bohm
gauge field requires a self-adjoint extension for the critical
subspace $n=k$. In fact, imposing regularity of both components of
the Dirac field at the origin is too strong a requirement, except
for integer flux. Rather, one has to apply the theory of Von
Neumann deficiency indices \cite{Reed}, which leads to a one
parameter family of allowed boundary conditions, characterized by
\cite{Gerbert}
\begin{equation}
i\,\lim_{r\rightarrow 0}\left(Mr\right)^{\nu +1}
g_n(r) \sin\left(\frac{\pi}{4}+\frac{\Theta}{2}\right)=
\lim_{r\rightarrow 0}\left(Mr\right)^{-\nu }
f_n(r) \cos\left(\frac{\pi}{4}+\frac{\Theta}{2}\right)
\end{equation}
with $\nu$ varying between $-1$ and $0$ ($\nu=-\alpha$ for $s=1$;
$\nu=\alpha-1$ for $s=-1$) . Here, $\Theta$ parametrizes the
admissible self-adjoint extensions, and $M$, a mass parameter, is
introduced for dimensional reasons.

Which of these boundary conditions to impose depends on the
physical situation under study.

One possibility is to take a finite flux tube, ask for continuity
of both components of the Dirac field at finite radius and then
let this radius go to zero \cite{AlRuWil1989,Hagen,Flekkoy}. Thus,
one of the possible self-adjoint extensions is obtained, which
corresponds to $\Theta= \frac{\pi}{2} {\rm
sgn}\left(\kappa\right)$.

As pointed out in \cite{AlRuWil1989,Sitenko1996gd,sitenko96}, this
kind of procedure leads to a boundary condition that breaks the
invariance under $\kappa\rightarrow\kappa+n$ ($n\,\epsilon\, Z$).
Now, this is a large gauge symmetry, which of course is singular
when considering the whole plane, but is not so when the origin is
removed or, equivalently, the plane has the topology of a
cylinder.

In order to preserve the aforementioned symmetry we propose,
instead, to exclude the origin, by imposing spectral boundary
conditions of the Atiyah-Patodi-Singer type \cite{aps,aps2,aps3},
as defined in \cite{Zhong}, at a finite radius $r_0$ and letting
$r_0\rightarrow0$.

We consider the development in eq.(\ref{eq-16}) and impose, at
$r=r_0$ :
\begin{equation}
\begin{array}{l}
  f_n\left(r_0\right) = 0\qquad{\rm for}\quad \lambda_n\left(r_0\right)\leq 0 \\
  g_n\left(r_0\right) = 0\qquad{\rm for}\quad \lambda_n\left(r_0\right)>0
\end{array}
\label{eq-19}
\end{equation}
for $s=1$.

As is well known, imposing this kind of boundary condition is
equivalent to removing the boundary, by attaching a semi-infinite
tube at its position and then extending the Dirac equation by a
constant extension of the gauge field, while asking that zero
modes be square integrable \cite{Ninomiya,Dowker6:1995} (except
for $\lambda_n=0$, where a constant zero mode remains, with a
nonzero lower component ).

Then, going to eq.(\ref{eq-17}) we have, after using the dominant
behaviour of Bessel functions for small arguments :
\begin{equation}
\begin{array}{l}
  \displaystyle{\frac{A_n}{B_n}}\sim r_0^{2\left(k+\alpha-n\right)}
  \qquad n-k+\frac{1}{2}-\alpha \leq 0 \\
  \displaystyle{\frac{B_n}{A_n}}\sim r_0^{2\left(n+1-k-\alpha\right)}
  \qquad n-k+\frac{1}{2}-\alpha> 0
\end{array}
\end{equation}

Now, we analyze two different situations:

If $\alpha \geq \frac{1}{2}$
\begin{equation}
\begin{array}{l}
  \displaystyle{\frac{A_n}{B_n}}\rightarrow_{r_0\rightarrow 0} 0
  \qquad {\rm for}\quad n\leq k \\
  \displaystyle{\frac{B_n}{A_n}}\rightarrow_{r_0\rightarrow 0} 0
  \qquad {\rm for}\quad n\geq k+1
\end{array}
\label{eq-21}
\end{equation}
and the eigenfunctions in eq(\ref{eq-18}) are of the
form:
\begin{equation}
\Psi_E\left(r,\theta\right) =
 \sum_{n=-\infty}^k B_n\left(\begin{array}{l}
   J_{k+\alpha-n}\left(|E|r\right) e^{i n \theta} \\
   -i\frac{|E|}{E}J_{k+\alpha-n-1}\left(|E|r\right) e^{i\left(n+1\right) \theta}
 \end{array}\right)+
 \sum_{n=k+1}^\infty A_n\left(\begin{array}{l}
   J_{n-k-\alpha}\left(|E|r\right) e^{i n \theta} \\
   i\frac{|E|}{E}J_{n+1-k-\alpha}\left(|E|r\right) e^{i\left(n+1\right) \theta}
 \end{array}\right)
\label{eq-22}
 \end{equation}

 If $\alpha<\frac{1}{2}$
\begin{equation}
\begin{array}{l}
  \displaystyle{\frac{A_n}{B_n}}\rightarrow_{r_0\rightarrow 0} 0
  \qquad {\rm for}\quad n\leq k-1 \\
  \displaystyle{\frac{B_n}{A_n}}\rightarrow_{r_0\rightarrow 0} 0
  \qquad {\rm for}\quad n\geq k
\end{array}
\label{eq-23}
\end{equation}
and the eigenfunctions are:
\begin{equation}
\Psi_E\left(r,\theta\right) =
 \sum_{n=-\infty}^{k-1} B_n\left(\begin{array}{l}
   J_{k+\alpha-n}\left(|E|r\right) e^{i n \theta} \\
   -i\frac{|E|}{E}J_{k+\alpha-n-1}\left(|E|r\right) e^{i\left(n+1\right) \theta}
 \end{array}\right)+
 \sum_{n=k}^\infty A_n\left(\begin{array}{l}
   J_{n-k-\alpha}\left(|E|r\right) e^{i n \theta} \\
   i\frac{|E|}{E}J_{n+1-k-\alpha}\left(|E|r\right) e^{i\left(n+1\right) \theta}
 \end{array}\right)
  \label{eq-24}
 \end{equation}

 Notice that our procedure leads precisely to a self adjoint
 extension satisfying the condition of minimal irregularity (the
 radial functions diverge at $r\rightarrow 0$ at most as $r^{-p}$,
 with $p\leq \frac{1}{2}$). It corresponds to the values of the
 parameter $\Theta$ :
\begin{equation}
 \Theta =\left\{\begin{array}{rl}
   -\frac{\pi}{2} & \quad{\rm for}\, \alpha\geq\frac{1}{2} \\
   \frac{\pi}{2} & \quad{\rm for}\, \alpha<\frac{1}{2} \

 \end{array}\right.
 \end{equation}

 As shown in \cite{manuel} $\Theta=\pm\frac{\pi}{2}$ are the
 only two possible values of the parameter which correspond to
 having a Dirac delta magnetic field at the origin.

 Moreover,
 this extension is compatible with periodicity in $\kappa$. In
 fact, the dependence on $k$ can be reduced to an overall phase
 factor in the eigenfunctions.

 As regards charge conjugation
 \[
 \Psi_E\rightarrow \sigma_1 {\Psi_E}^*\quad;\quad
 \kappa\rightarrow-\kappa
 \]
 it is respected by the eigenfunctions, except for
 $\alpha=\frac{1}{2}$. This is due to the already commented
 presence of a constant zero mode on the cylinder.

Notice that, for the representation $s=-1$ of $2\times 2$ Dirac
matrices,
\begin{equation}
\Psi_E^{(-)}=\left(\begin{array}{l}
  e^{i \frac{\theta}{2}} \sum f_n e^{i \left(n+\frac{1}{2}\right)\theta} \\
  e^{-i \frac{\theta}{2}} \sum g_n e^{i\left(n+\frac{1}{2}\right)\theta}
\end{array}
\right)
\end{equation}
and APS boundary conditions must be reversed, for $\lambda\neq0$,
as compared to (\ref{eq-19}), since the operator ${\rm B}$ changes
into $-{\rm B}$. For $\lambda=0$ the lower component will be taken
to be zero at $r_0$ which, as we will show later, allows for
charge conjugation to be a symmetry of the whole model. Thus we
take
\begin{equation}
\begin{array}{l}
  f_n\left(r_0\right) = 0\qquad{\rm for}\quad \lambda_n\left(r_0\right)>0 \\
  g_n\left(r_0\right) = 0\qquad{\rm for}\quad \lambda_n\left(r_0\right)\leq 0
\end{array}
\label{eq-19p}
\end{equation}
for $s=-1$

In this case, the resulting extension corresponds to
\begin{equation}
 \Theta =\left\{\begin{array}{rl}
   \frac{\pi}{2} & \quad{\rm for}\, \alpha\geq\frac{1}{2} \\
   -\frac{\pi}{2} & \quad{\rm for}\, \alpha<\frac{1}{2} \

 \end{array}\right.
 \end{equation}

 This is an example of a physical application of APS boundary conditions,
 which are generally chosen due to their mathematical interest. We
 wish to stress that our conclusions concerning the behaviour at
 the origin hold true for massive Dirac fields. The procedure of
 imposing boundary conditions at a finite radius, then taken to
 zero, was considered for massive fields in \cite{Poly}, where
 some vacuum quantum numbers were examined.

 It is worth pointing that, for integer $\kappa=k$, our procedure
 leads (both for $s=\pm1$) to the requirement of regularity of both
 components at the origin. In this case
\begin{equation}
\Psi_E\left(r,\theta\right) =  \sum_{n=-\infty}^\infty A_n
\left(\begin{array}{l}
  J_{n-\kappa}\left(|E|r\right) e^{i n\theta} \\
  i\frac{|E|}{E} J_{n+1-\kappa}\left(|E|r\right)e^{i \left(n+1\right)\theta}
\end{array}
\right)
\end{equation}

\section{The theory in a bounded region. Zero modes and APS index
theorem}
\label{section-4}

From now on, we will confine the Dirac fields inside a bounded
region, by introducing a boundary at $r=R$, and imposing there
boundary conditions of the APS type, complementary to the ones
considered at $r=r_0$.

For $s=1$, these boundary conditions are
\begin{equation}
\begin{array}{c}
  f_n\left(R\right) = 0\qquad{\rm for}\quad \lambda_n\left(R\right)>0 \\
  g_n\left(R\right) = 0\qquad{\rm for}\quad \lambda_n\left(R\right)\leq 0
\end{array}
\label{eq-25}
\end{equation}

We start by studying the zero modes of our theory, which are of
the form
\begin{equation}
\Psi_0\left(r,\theta\right)=\left(\begin{array}{l}
  e^{-i \frac{\theta}{2}}
  \sum_{n=-\infty}^\infty  A_n r^{n-k-\alpha}
  e^{i \left(n+\frac{1}{2}\right)\theta} \\
  e^{i \frac{\theta}{2}}  \sum_{n=-\infty}^\infty  B_n r^{k+\alpha-n-1}
  e^{i\left(n+\frac{1}{2}\right)\theta}
\end{array}
\right) \label{eq-26}
\end{equation}

The conditions at $r=r_0$ imply
\begin{equation}
\begin{array}{c}
  A_n = 0 \qquad,\quad{\rm for}\quad n\leq k+\alpha-\frac{1}{2} \\
  B_n = 0 \qquad,\quad{\rm for}\quad n> k+\alpha-\frac{1}{2}
\end{array}
\end{equation}

Now, the boundary conditions at $R$ imply
\begin{equation}
\begin{array}{c}
  A_n = 0 \qquad,\quad{\rm for}\quad n> k+\alpha-\frac{1}{2} \\
  B_n = 0 \qquad,\quad{\rm for}\quad n\leq k+\alpha-\frac{1}{2}
\end{array}\label{eq-28}
\end{equation}

Thus, no zero mode remains under this boundary condition, even
without taking $r_0\rightarrow 0$. This is in agreement with the
APS index theorem \cite{aps,aps2,aps3}. In fact, according to such
theorem
\begin{equation}
n_+ -n_- = {\cal A} + b\left(r_0\right)+b\left(R\right)
\end{equation}
where $n_+ (n_-)$ is the number of chirality positive (negative)
zero energy solutions, ${\cal A}$ is the anomaly, or bulk
contribution, and $b$ are the surface contributions coming from
both boundaries \cite{spectral,Zhong}
\begin{equation}
b\left(R\right) = \frac{1}{2}\left(h_R-\eta \left(R\right)\right)
\qquad b\left(r_0\right) =
\frac{1}{2}\left(\eta\left(r_0\right)-h_{r_0}\right)
\end{equation}
with
\begin{equation}
\eta\left(r\right) = \left.\sum_{\lambda_n\left(r\right)\neq 0}
{\rm sgn}
\lambda_n\left(r\right)\,|\lambda_n\left(r\right)|^{-s}\right\rfloor_{s
=0}
\end{equation}
the spectral asymmetry of the boundary operator ${\rm B}$ and
$h_r$ is the dimension of its kernel.

In our case :
\begin{equation}
b\left(r_0\right) =\left\{\begin{array}{ll}
  \alpha -1 & \quad \alpha>\frac{1}{2} \\
  -\frac{1}{2} & \quad \alpha = \frac{1}{2} \\
  \alpha & \quad \alpha<\frac{1}{2}
\end{array}\right.
\qquad b\left(R\right) =\left\{\begin{array}{ll}
  1-\alpha  & \quad \alpha>\frac{1}{2} \\
  \frac{1}{2} & \quad \alpha = \frac{1}{2} \\
  -\alpha & \quad \alpha<\frac{1}{2}
\end{array}\right.
\end{equation}

Thus, the boundary contributions cancel. As regards the volume
part, it also vanishes for the gauge field configuration under
study, and we have $n_+ -n_- =0$, which is consistent with our
analysis following eq.(\ref{eq-28}).

For $s=-1$ both boundary contributions interchange, and identical
conclusions hold regarding the index.

\section{Energy spectrum. Fermionic number and Casimir energy}
\label{section-5}

The energy spectrum can be determined by imposing (for $s=1$) the
boundary conditions (\ref{eq-25}) at $r=R$ on the eigenfunctions
of eq.(\ref{eq-22}) if $\alpha\geq \frac{1}{2}$, or on the
eigenfunctions in eq.(\ref{eq-24}), if $\alpha<\frac{1}{2}$. When
doing so, we obtain :
\begin{equation}
E_{n,l}=\left\{
\begin{array}{l}
  \pm\frac{j_{n-\alpha,l}}{R} \qquad ,
  \quad n=1,\ldots \infty \quad l=1 \ldots \infty\\
  \pm\frac{j_{n+\alpha,l}}{R} \qquad ,
  \quad n=-1,\ldots \infty \quad l=1 \ldots \infty
\end{array}
\right. \qquad {\rm for}\,\alpha\geq \frac{1}{2} \label{eq-29}
\end{equation}
and
\begin{equation}
E_{n,l}=
\left\{
\begin{array}{l}
  \pm\frac{j_{n-\alpha,l}}{R} \qquad ,
  \quad n=0,\ldots\infty \quad l=1\ldots\infty\\
  \pm\frac{j_{n+\alpha,l}}{R} \qquad ,
  \quad n=0,\ldots\infty \quad l=1\ldots\infty
\end{array}
\right. \qquad {\rm for}\,\alpha< \frac{1}{2} \label{eq-30}
\end{equation}
where $j_{\nu,l}$ is the $l$-th positive root of $J_\nu$. The same
spectrum results for $s=-1$.

For both $s$ values, the energy spectrum is symmetric with respect
to zero. This fact, together with the absence of zero modes
results in a null vacuum expectation value for the fermionic
charge \cite{Niemi1984} :
\begin{equation}
<N>_{+} =-\frac{1}{2}\left(n_+-n_-\right)= 0
\end{equation}

For the same reasons $<N>_{-} =0$, so the total fermionic number
of the theory is null.

It is interesting to note that the origin contributes to the
fermionic number with (see the discussion in \cite{Poly})
\begin{equation}
<N>_{r_0,\pm}=
\left\{\begin{array}{ll}
  \mp\frac{1}{2}\left(\alpha -1\right) & \quad \alpha>\frac{1}{2} \\
  \pm\frac{1}{4} & \quad \alpha = \frac{1}{2} \\
  \mp\frac{1}{2}\alpha & \quad \alpha<\frac{1}{2}
\end{array}\right.
\end{equation}
which coincides, for each $s$ value, with the result presented for
the whole punctured plane in \cite{Sitenko1996gd,sitenko96}, where
opposite signs of the mass correspond to our opposite signs of $s$
(see also \cite{Moroz} for related work) , except for
$\alpha=\frac{1}{2}$. This last fact is associated with charge
conjugation non invariance in each subspace. However, the sum of
both contributions cancels for all $\alpha$.
\bigskip

We go now to the evaluation of the Casimir energy, which is
formally given by

\begin{equation}
E_C = -\frac{1}{2}
\left( \sum_{E>0} E_{n,l} -
\sum_{E<0} E_{n,l}
\right) = -\left(
\sum_{E>0} E_{n,l}
\right)\qquad,
\end{equation}
where the symmetry of the energy spectrum for our problem has been
used. Of course, a regularization method must be introduced in
order to give sense to this divergent sum. In the framework of the
$\zeta$-regularization \cite{zeta,Dowker:1976tf},
\begin{equation}
E_C =-\mu \left.\sum_{E>0} \left(\frac{E_{n,l}}{\mu}\right)^{-z}
\right\rfloor_{z=-1} \label{eq-33}
\end{equation}
where the parameter $\mu$ was introduced for dimensional reasons.

Here, it is useful to define the so-called partial zeta function
as in references \cite{Leseduarte1996,Romeo1993,Romeo1994}

\begin{equation}
\zeta_\nu\left(z\right) =\sum_{l=1}^\infty
\left(j_{\nu,l}\right)^{-z}\label{eq-34}
\end{equation}

So, for the problem at hand we have, from
eqs.(\ref{eq-29}),(\ref{eq-30}),(\ref{eq-33}) and (\ref{eq-34})
\begin{equation}
E_C =-2\mu\left. \left(\mu R\right)^z \sum_\nu
\zeta_\nu\left(z\right)\right\rfloor_{z=-1} \label{eq-35}
\end{equation}
with
\begin{equation}
\sum_\nu \zeta_\nu = \left\{
\begin{array}{ll}
  \sum_{n=-\infty}^\infty \zeta_{|n-\alpha|} \,+ \zeta_{\alpha-1} &
  \qquad {\rm for}\quad \alpha \geq\frac{1}{2} \\
  \sum_{n=-\infty}^\infty \zeta_{|n-\alpha|} \,+ \zeta_{-\alpha} &
  \qquad {\rm for}\quad \alpha <\frac{1}{2}
\end{array}
\right.
\label{eq-36}
\end{equation}

Notice that, as a consequence of the invariance properties of the
imposed boundary conditions, the Casimir energy is both periodic
in $\kappa$ and invariant under $\alpha\rightarrow 1-\alpha$, as
well as continuous at integer values of $\kappa$.

For any value of $\kappa$, it is the sum of the energy
corresponding to a scalar field in the presence of a flux string
and subject to Dirichlet boundary conditions (studied in
ref.\cite{Leseduarte1996}), plus a partial zeta coming, for
fractionary $\kappa$, from the presence of an eigenfunction which
is singular at the origin or, for integer $\kappa$, from the
duplication of $J_0$.

Both contributions can be studied following the methods employed
in \cite{Leseduarte1996} and developed by the same authors in
previous work \cite{Romeo1993,Romeo1994}. So, here we won't go
into the details of such calculation.

The scalar field contribution presents a pole at $z=-1$, with an
$\alpha$-independent residue. So, in this case, a renormalized
Casimir energy can be defined as
\begin{equation}
E_C^{ren} =E_C\left(\alpha\right) -E_C\left(0\right) \label{eq-37}
\end{equation}

Following \cite{Leseduarte1996}, this quantity can be obtained
through a numerical calculation. (Notice there is a minor error in
eq.(3.6) of that reference. In fact, a factor of $\pi$ is missing
in the fifth term on the right hand side of such equation).

Now, going back to the Dirac field, the partial zeta appearing in
(\ref{eq-35}),(\ref{eq-36}) can also be evaluated with the methods
of \cite{Leseduarte1996}. However, it presents a pole, whose
residue is $\alpha$-dependent :
\begin{equation}
\left. \zeta_\nu \left( z \right) \right\rfloor_{z=-1} =\left.
F.P. + \frac{1}{8\pi}\left( 1-4\nu^2\right)\frac{1}{z+1}
\right\rfloor_{z=-1}
\end{equation}

Notice that such residue vanishes only for $\nu=\pm\frac{1}{2}$
($\alpha=\frac{1}{2}$ in the case of interest).

\begin{figure}
\epsffile{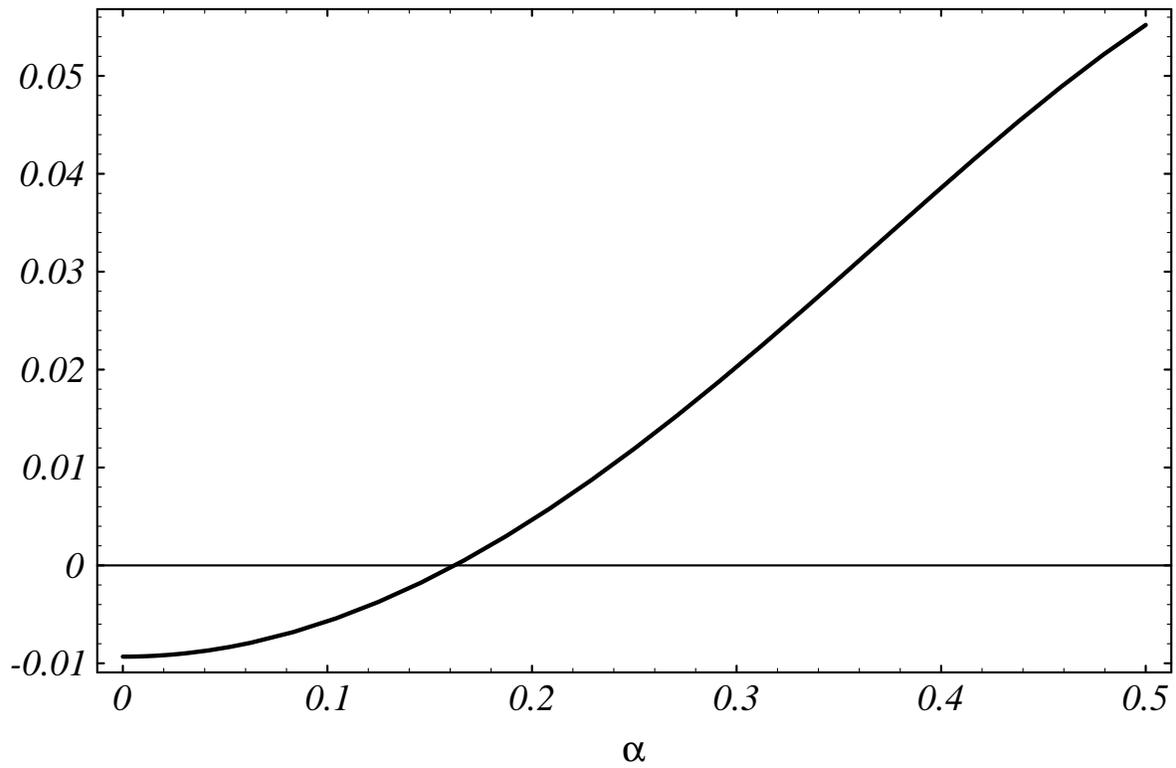} \caption{Finite part of the adimensionalized
Casimir Energy} \label{fig-1}
\end{figure}

The finite part of the Casimir energy is plotted in
fig.(\ref{fig-1}) as a function of $\alpha$ for $\mu R=1$. Of
course, an absolute meaning cannot be assigned to it, due to the
presence of the pole and the consequent need to introduce
$\alpha$-dependent counterterms.

\section{Comments and conclusions}
\label{section-6}

In summary, we have studied the problem of Dirac fields in the
presence of a Bohm-Aharonov background which can, by virtue of the
symmetry properties of the external field, be separated into two
2+1-dimensional ones (labelled through the paper by $s=\pm 1$).

We have adopted the viewpoint that the plane is a punctured one,
so that the model must turn out to be invariant under integer
translations of the reduced flux, which are nothing but
``singular" gauge transformations. In this spirit we have shown
that, by imposing spectral boundary conditions of the APS type at
a finite radius and then letting this radius to zero, one among
the family of allowed boundary conditions at the origin is
obtained for the two-component spinors. Under this boundary
condition, the dependence on the integer part of the flux can be
extracted from the eigenfunctions of the Hamiltonian as an overall
phase. As a consequence, the aforementioned symmetry is preserved.

As regards the invariance under charge conjugation, it is broken
for each $s$ value by the boundary conditions at the origin, for
the fractionary part of the flux equal to $\frac{1}{2}$. This is
due to the fact that a constant zero mode of definite handedness
is allowed for on the cylinder when applying the APS boundary
conditions. However, this symmetry is preserved when opposite
handed conditions are imposed on both $s$-subspaces.

It is worth stressing that this approach to the problem of
self-adjointness at the origin is, to our knowledge, the first
proposal of a physical application of APS boundary conditions in
this context.

We have then confined the theory to a finite region, i.e. a
punctured circle, and imposed adequate APS boundary conditions at
the exterior boundary, with the aim of checking consistency  with
the APS index theorem for manifolds with boundaries, which we have
shown to hold.

Finally, we have studied the vacuum fermionic number and analized,
through zeta function regularization techniques, the Casimir
energy of the model. Since this last turns out to present a
divergence depending on the fractionary part of the flux, no
absolute significance can be assigned to its finite part, which is
thus subject to renormalizations.

However, notice that, contrary to the treatment of the origin, APS
boundary conditions were imposed at the exterior boundary for
merely formal reasons (the existence of an index theorem for this
case). The vacuum energy under local (bag-like) external boundary
conditions is at present under study.

\acknowledgments

The authors thank H. Falomir for many useful discussions and
comments.

\bigskip

This work was partially supported by ANPCyT (Agencia Nacional de
Promoci\'{o}n Cient\'{\i}fica y Tecnol\'{o}gica), under grant PMT-PICT
0421 and U.N.L.P. (Universidad Nacional de La Plata), Argentina.



\bigskip

Paper: Dirac fields in the background of a magnetic flux string
and spectral boundary conditions

Authors: C. G. Beneventano, M. De Francia and E. M. Santangelo

\pagebreak

\section*{Figure Caption}

\bigskip

FIG. 1 - Finite part of the adimensionalized Casimir Energy

\end{document}